\documentclass{emulateapj}             
             
\shorttitle{Radiation form cocoon} \shortauthors{Pe'er, Meszaros \& Rees}

\newcommand{\eV}{\rm{\, eV }}        
\newcommand{\keV}{\rm{\, keV }}

\newcommand{\beq}{\begin{equation}}             
\newcommand{\eeq}{\end{equation}}             
\newcommand{\ba}{\begin{array}}             
\newcommand{\ea}{\end{array}}

\newcommand{\Gi}{\Gamma_{2}}          
\newcommand{\Gc}{\Gamma_{c,1}}

\def \etal{{\it et al.~}}            
            
\begin{document}             
          
\title{Radiation from an expanding cocoon as an explanation of the steep      
decay observed in GRB early afterglow light curves}         
          
\author{Asaf Pe'er\altaffilmark{1}\altaffilmark{2},      
 Peter M\'esz\'aros\altaffilmark{2} and Martin J. Rees\altaffilmark{3} }         
\altaffiltext{1}{Astronomical Institute ``Anton Pannekoek'', Kruislaan 403, 1098      
SJ Amsterdam, the Netherlands; apeer@science.uva.nl}         
\altaffiltext{2}{Dpt. Astron. \& Astrophysics, Dpt. Physics, Pennsylvania State       
University, University Park, PA 16802}          
\altaffiltext{3}{Institute of Astronomy, University of Cambridge, Madingley Rd.,      
 Cambridge CB3 0HA, UK}

\begin{abstract}      
Observations of early afterglow emission from gamma-ray bursts (GRB's)  
with the $Swift$ satellite show steep decay of the X-ray light curve,  
$F_\nu(t) \propto t^{-\alpha}$ with $\alpha \approx 2.5 - 4$ at $\sim  
300 - 500$~s after the burst trigger. The spectrum in this time  
interval is consistent with a spectrum $F_\nu \propto \nu^{-\beta}$  
with $\beta \approx 1$.  Here, we show that these results can be  
explained as due to emission from the hot plasma ``cocoon" associated  
with the jet, which expands relativistically after the jet has broken  
through the stellar envelope, if a substantial fraction of the cocoon  
kinetic energy is dissipated at scattering optical depths $\tau_T\sim  
10^2-10^3$. This results in the bulk of the cocoon photons being  
observed at X-ray energies, after a delay of few hundreds of seconds  
relative to the gamma-ray photons from the jet.  Multiple Compton  
scattering inside the cocoon causes a spread in the arrival times of  
the X-ray photons.  We calculate numerically the observed light curve  
of photons emerging from the cocoon, and show that it exhibits a steep  
decay, which resembles that observed in many GRB afterglows.  During  
the adiabatic expansion that follows the dissipation phase, photons  
lose energy to the expanding plasma, and as a result, the emerging  
photon energy distribution differs from a black-body spectrum, and can  
be approximated as a power law in the $Swift$ XRT band.  Comparison of  
the numerical results with the $Swift$ XRT data of GRB050315 and  
GRB050421 shows good agreement between the light curves and spectra  
during the initial steep decay phase.  
     
\end{abstract}          
      

\section{Introduction}      
\label{sec:intro}      
      
The successful launch of the $Swift$ gamma-ray bursts (GRB) explorer      
\citep[e.g.,][]{Gehrels04} enabled to probe X-ray afterglow emission      
at early times ($\sim 10^2$~s to $\sim 10^4$~s) after the burst onset,      
a time interval which was largely unexplored previously.  The early      
afterglow observations revealed an unexpectedly steep decline in the      
X-ray light curves of $\sim 2/3$ of the bursts that did not show      
flaring activity, $F_\nu(t) \propto t^{-\alpha}$ with $\alpha \approx      
2.5 - 4$ following the prompt emission and lasting few tenth -      
hundreds of seconds. The steep decline in the light curve is followed      
by a much shallower decay, $\alpha \approx 0.7$ at later times      
\citep{Tagliaferri05, Chincarini05, Zhang05, Nousek05, Obrien05}.  The      
steep decay segment observed by the X-ray telescope      
\citep[XRT;][]{Burrows05} usually connects to the spectral      
extrapolation of the Burst Alert Telescope      
\citep[BAT;][]{Barthelmy05a} prompt emission light curve smoothly      
\citep{Cusumano05, Barthelmy05, Hill05}.      
      
The spectral index obtained by the XRT in the 0.3-10 \keV range during      
the steep decay segment of the light curve is close to unity, $F_\nu      
\propto \nu^{-\beta}$, with $\beta \simeq 1 $ \citep{Obrien05}. This      
power law index is somewhat softer than the power law index obtained      
at higher energies by the BAT data, at 15-150 \keV.          
      
A possible explanation to the steep decline observed in many light      
curves was suggested by \citet{Zhang05} and \citet{Nousek05}      
\citep[see also ][]{Liang06}.       
According to their suggestion,      
the steep decline is due to the effect known as ``the curvature      
effect'' \citep{Fenimore96, KP00, Der04}.      
In this effect, radiation arising from high angular latitude relative      
to the viewing direction arrives to the observer at late times,      
following the termination of the prompt emission, due to the extra       
distance it travels. In addition, this radiation is observed at flux      
and peak frequency lower than the corresponding in the prompt      
emission radiation, due to the beaming effect.       
This model predicts a connection between the temporal and spectral      
indices of the emission \citep{KP00, Der04, Zhang05}: in the relation      
$f_\nu(t) \propto t^{-\alpha} \nu^{-\beta}$, $\alpha$ and $\beta$ are      
related via $\alpha = 2 + \beta$.        
However, a detailed analysis of 40 XRT light curves by      
\citet{Obrien05} found that the spectral and temporal indices are       
 uncorrelated, in contrast to this model prediction. Moreover, when      
 fitting the light curves in a model that allows stretching and shifting      
 of the trigger time, \citet{Obrien05} found that 6 out of the 40      
 bursts show spectral decay       
 index larger than the allowed value of the high latitude      
 emission model prediction.        
      
      
      
In recent years, there is increasing evidence that long duration      
\citep[$t_{90}\geq$ 2 s;][]{K93} GRB's are associated with the deaths      
of massive stars, presumably arising from core collapse      
\citep{Woosley93, LE93, MW99, MWH01, ZWM03} \citep[see also][for short      
  summary of observational evidence]{PW05}.      
In this so called ``collapsar'' model \citep{Woosley93, Pac98},      
the collapse of the Fe core of a massive star leads to the formation      
of a central black hole (BH) of several solar masses. A relativistic      
jet is thought to arise during this short accretion period      
\citep{Aloy00, MWH01}. As the light (compared to the stellar density),      
relativistic jet makes its way out of the progenitor star, its rate of      
advance is slowed down, and most of the energy output during that      
period is deposited into a hot cocoon surrounding it \citep{MR01,      
  Matz03, RCR02}.       
      
The jet head propagates through the Fe and He core of the star, until      
it emerges from the He core into the low density H envelope at      
$r_\star \approx 10^{11}$~cm \citep{MW99, MR01, Matz03, WM03}.       
Following the jet emergence, the hot plasma composing the cocoon      
(shocked jet material which accumulates while the jet forces its way      
out, entrained with material swept from the star)       
swiftly escapes from the stellar cavity and accelerates in      
approximately the same way as the plasma composing the jet - its      
energy is converted via adiabatic expansion into bulk kinetic energy      
\citep{MR01, RCR02}.       
The entropy per baryon in the cocoon, $\eta_c = (L_c/\dot{M} c^2)$, is      
smaller than the entropy per baryon in the jet, due to the entrainment      
of the cocoon with material from the star. The cocoon therefore      
accelerates up to an asymptotic Lorentz factor which is lower than the      
Lorentz factor of the jet (but still, relativistic), before      
dissipating its kinetic energy.      
      
The larger number of baryons in the cocoon compared to the jet baryon      
load implies that if the cocoon kinetic energy is dissipated at      
similar radii to the dissipation radius of the jet kinetic energy      
(e.g., by magnetic reconnection, as suggested by \citet{Thompson94}      
and \citet{GS05}, or alternatively by shock waves), than the optical      
depth for scattering of photons inside the       
cocoon at the dissipation radius is very large, $\tau_{\gamma e} \sim      
10^2 - 10^3$. As a result, these photons cannot emerge from the cocoon       
immediately, and are diffused out during the adiabatic expansion that      
follows the dissipation. These photons thus suffer a significant time      
delay compared to photons that were emitted from the jet as the jet      
kinetic energy was dissipated, which produce the observed prompt emission.      
Moreover, the multiple scattering of these photons inside the cocoon      
produces a spread in the escape times of these photons, which - as we      
show below - can result in the steep slopes in the early afterglow      
light curves observed by the {\it Swift}.       
      
In this paper, we analyze the emission from the expanding cocoon, and      
show that it can provide a natural explanation to the observed decline in      
the early afterglow light curves.  We calculate in      
\S\ref{sec:time_delay} the diffusion time delay of the cocoon      
photons. In \S\ref{sec:light_curve} we present numerical results of      
the light curves and spectra. We show that the resulting spectra is      
much flatter than black-body spectrum, due to the multiple scattering      
by the cocoon plasma electrons, during the cooling down of the      
electrons in the adiabatic expansion that follows the dissipation.  We      
compare our results with the {\it Swift} data in \S\ref{sec:data},      
before summarizing in \S\ref{sec:summary}.

\section{Scaling laws for the time delay and temperature of cocoon photons}      
\label{sec:time_delay}      
      
Following the emergence of the relativistic jet from the star (e.g. a  
WR star), the hot plasma composing the cocoon escapes from the jet  
cavity. As pointed out by \citet{LazBeg05}, the jet opening angle at  
breakout is $\lesssim 1^\circ$, while the cocoon opening angle is of  
the order of tens of degrees. Following its emergence, the cocoon  
expands in all directions in the dilute stellar wind or dilute  
H-envelope (if any). At this stage, the lateral expansion  
velocity of the cocoon is close to the speed of light, and the cocoon  
expansion becomes isotropic \citep[e.g., ][]{Aloy00}. We can therefore  
approximate the cocoon geometry in its expansion phase as spherical. 
   
During the cocoon expansion, which lasts few - few tens of seconds (see    
below), the inner engine that produces the jet may still remain active.   
Therefore, during the cocoon expansion, fresh jet material continuously    
passes through the cocoon, in a similar way to jet propagation in the core    
of the star, thus keeping the cocoon channel open. During its expansion    
the cocoon pressure decreases, therefore the jet material passing through the   
cocoon is less tightly collimated, resulting in an increase of the jet    
opening angle as it passes through the cocoon \citep[e.g., see discussion    
in][]{LazBeg05}.    
   
Dissipation of the jet kinetic energy produces the prompt $\gamma$-ray   
emission. This dissipation can be caused by e.g., internal collisions or 
magnetic reconnection occurring at radii larger than that of the    
leading edge of the cocoon (which is less relativistic than the jet,    
$\Gamma_{co} \sim 10$ vs. $\Gamma_{jet}\gtrsim 100$). Thus, the cocoon   
does not interfere with the usual jet prompt emission production.    
The presence of features in the early X-ray light curve such as X-ray    
bumps or flares, or a shallow X-ray decay phase, should also be compatible   
with the presence of a cocoon. While specific mechanisms for flares and    
shallow decays are not considered here, if these are produced, e.g. by   
late internal shocks or continued injection of energy into the jet,    
a requirement is that the funnel drilled by the jet inside the cocoon    
remains open during the prompt and early afterglow emission  
phase. This is plausible    
e.g. in a late internal shock or continued ejection model, where material   
continues to flow out in the jet. This in turn, suggests that disturbances   
and oblique shocks in the expanding cocoon, caused by the continued jet    
activity, can lead to additional dissipation throughout the cocoon volume,   
in addition to the cocoon interaction with the external medium or wind.   
Photons emitted in the jet, whether prompt $\gamma$-rays, early flares or   
shallow decay, are expected to escape immediately, provided as usual that   
they are emitted above the jet photosphere. They are also beamed into a solid   
angle smaller than the jet opening angle. However, photons emitted in the    
cocoon surrounding the jet cannot escape directly, due to the high optical    
depth in the cocoon.   
   
The cocoon photons are therefore observed with a time delay compared to    
the prompt $\gamma$-ray photons emitted from the jet (although they can   
arrive before the photons produced by late internal shocks or refreshed   
shocks in the jet which might produce flares and shallow decays).   
This time delay relative to the prompt $\gamma$-rays has three  
different sources.     
First, $\Delta t_g$ is the ``geometrical'' time delay, representing      
the time delay of photons originating from different regions in the      
flow compared to photons originating from the line of sight toward the      
observer.  Second, $\Delta t_{esc}$ is the time delay resulting from the      
finite time it takes the cocoon plasma to emerge from the Fe/He core of      
the star, following the emergence of the jet. Third, $\Delta t_D$ is      
the diffusion time delay, which is the time it takes the cocoon      
photons to escape from the cocoon plasma.

In order to estimate these time delays, we assume that both the cocoon  
and the jet emerge from the core of the star at $r_\star = 10^{11}  
r_{\star,11}$~cm \citep{MR01,WM03,Matz03,LazBeg05}.  After the jet  
drills its way through the core of the star, it propagates through the  
H envelope, if any, and escapes, producing the ``prompt" $\gamma$-ray  
emission at some usual nominal radius $r_i\simeq 10^{13} r_{i,13}$ cm,  
where the jet is already optically thin.  Relative to some nominal  
signal emitted from the base of the flow, these photons reach the  
observer subject to a geometrical time delay $\Delta t_{g,jet} \simeq  
r_i/(4 \Gamma_j^2 c) \simeq 8 \, r_{i,13} \Gi^{-2} \, $~ms, due to the  
finite width and angular size of the jet emitting region.  
     
The cocoon starts its acceleration once it emerges from the core of    
the star, when the jet has broken through. We approximate the time it    
takes the cocoon plasma to emerge from the core as $\Delta t_{esc} = f    
r_\star / c_s \approx 6 \, r_{\star,11} \, f_0$~s, where $c_s = c/\sqrt{3}$    
is the speed of sound and $f \geq 1$ is an unknown numerical factor    
which accounts for the excess of time (above the free expansion time)    
it takes the cocoon to emerge from the core, and is taken here to be    
$f = 1 f_0$.  To estimate the total energy in the cocoon, we start from    
the Frail et al (2001) value for the typical (collimation corrected)    
jet energy, $E_j\sim 10^{51}$ erg, with a spread of about one order of    
magnitude.  This refers mainly to the electron component, so    
conservatively one can multiply by a factor 3 for the proton energy in    
the jet. The cocoon is fed by the total output of the jet while the    
latter expands subrelativistically inside the star, and assuming a    
factor $\sim 3$ longer jet lifetime inside the star than outside it,    
the cocoon total energy may be taken as $E_c\sim 10^{52}E_{c,52}$    
erg. The baryon mass in the cocoon is also expected to be significantly    
larger by an uncertain value than that in the jet, due to entrainment    
of baryons from the stellar envelope. The entropy per baryon in the    
cocoon plasma is likely to be lower than that of the jet, and is    
parametrized here as $\eta_c = 10 \eta_{c,1}$. Following its emergence    
from the core of the star, the cocoon expands spherically and    
accelerates up to a cocoon saturation radius, $r_{s,c} = r_\star    
\eta_c = 10^{12} \, r_{\star,11} \Gamma_{c,1} $~cm, where $\Gamma_c =    
\eta_c = 10 \Gamma_{c,1}$ is the asymptotic value of the cocoon    
Lorentz factor \citep[e.g.,][]{RCR02}.   
Assuming isotropic expansion of the cocoon outside the core, the    
comoving cocoon width is estimated by $\Delta r_c \approx \Gamma_c    
r_\star = 10^{12} \, r_{\star,11} \Gamma_{c,1}$~cm.    
     
The cocoon energy release can be approximated as an impulsive release    
involving a single pulse of width $\Delta t_{esc}$. Thus, at a radius    
$r_{i,c} > r_{s,c}$ where dissipation of the cocoon kinetic energy    
occurs (e.g. via internal shocks, reconnection, etc), the cocoon    
comoving proton density can be estimated as $n_p(c) = E_c/(4 \pi    
r_{i,c}^3 m_p c^2)$, where $E_c$ is the energy content of the    
cocoon. For dissipation at a radius $r_{i,c}$ below the cocoon    
photospheric radius where the cocoon becomes optically thin, the    
cocoon optical depth to scattering is given by    
\beq       
\tau_{\gamma e,c} = \Delta r_c n_p(c)      
\sigma_T \simeq 350 \, E_{c,52} r_{\star,11} r_{i,13}^{-3}    
\Gamma_{c,1},     
\label{eq:tau}     
\eeq       
where $E_c = 10^{52}E_{c,52}$~erg. Thus, for dissipation radii anywhere     
between the stellar radius $r_{\star,11}$ and the cocoon photosphere at     
$r\sim 10^{14}$~cm, the optical depth to scattering inside the cocoon is     
high.  Therefore, photons that are produced inside the cocoon suffer,     
in addition to a geometrical time delay as they emerge from the photosphere,      
$\Delta t_{g,c} = r_{ph}/4\Gamma_c^2 c $, also a diffusive time delay before      
they escape.      
     
     
The diffusive time delay is estimated as follows. Following the      
dissipation phase, the cocoon expands in its rest frame at the speed      
of sound.  The electron temperature inside the cocoon is mildly      
relativistic, $kT'/m_e c^2 \simeq 10^{-3}$ (see below), and the      
cocoon is radiatively dominated, therefore the cocoon expansion speed      
is close to the asymptotic value $v \lesssim c/\sqrt{3}$.      
The rapid expansion implies that the mean free path of      
photons inside the cocoon increases with (comoving) time as       
$l(t) \simeq l_0 (v t/\Delta r_c)^3$, where       
$l_0 = \Delta r_c /\tau_{\gamma e,c}$       
is the mean free path at the end of the dissipation phase, and      
a constant expansion velocity $v$ assumed.       
For a large number of scatterings, $\tau_{\gamma e,c} \gg 1$, the      
distance traveled by a photon before it escapes is      
$c t = \int (dl(t)/dt) dt$, where $t$ is the comoving time from photon      
production to escape. We thus find that the comoving diffusion time is      
\beq      
t \equiv \Delta t_D^{co.} \simeq {\sqrt{\tau_{\gamma e,c}} \Delta r_c      
  \over c},      
\eeq       
and the observed diffusive time is shorter by a factor of $\Gamma$.      
A full analytical treatment presented in \S\ref{sec:appendix}      
gives a numerical pre-factor $3^{3/4}$ (for $v=c_s$).      
An alternative calculation of the time delay can be carried out using the      
fact that the photons emerge from the expanding plasma as they reach the     
photospheric radius, $r_{ph}$. We show in \S\ref{sec:appendix2} that      
the two approaches yield similar results.       
Nonetheless, the advantage of the calculation as presented here is      
that it can be generalized to the calculation of       
the spread in the photon arrival time, which results in the steep      
decay (see \S\ref{sec:light_curve} below).

The comoving temperature of the cocoon plasma at the dissipation      
radius is calculated as follows. During the cocoon emergence time      
from the star $\Delta t_{esc}$, the cocoon expands to a radius $\sim      
\Delta t_{esc} \times c_s$, thus the cocoon occupies a volume $V_c =      
(4\pi/3) (f r_\star)^3$.  As it emerges from the core of the star, the     
cocoon temperature $T'(r_c)$ is calculated using $E_c/V_c \simeq a      
T'(r_c)^4$, where $a$ is Stefan's constant.  At $r<r_{s,c}$, $T'(r)       
\propto r^{-1}$, and at larger radii. $T'(r) \propto r^{-2/3}$. We      
thus find that at the dissipation radius,      
\beq       
T'(r_i) = \left(3 E_c \over 4 \pi (f r_\star)^3 a \right)^{1/4}       
\left({r_{s,c} \over r_\star}\right)^{-1} \left({r_i \over      
  r_{s,c}}\right)^{-2/3}.       
\label{eq:temp0}      
\eeq       
The normalized cocoon temperature at the dissipation radius is      
therefore equal to       
\beq       
\theta_{el} \equiv {k_B T' \over m_e c^2} = 5 \times 10^{-4}\,      
E_{c,52}^{1/4} \, r_{\star,11}^{-1/12} \, r_{i,13}^{-2/3} \, \Gc^{-1/3} \, f_0^{-3/4}.       
\label{eq:temp}      
\eeq

\section{Light Curve and spectra of photons emerging from the cocoon}      
\label{sec:light_curve}      
      
While the details of the dissipation process are uncertain, the high      
optical depth inside the cocoon implies that photons undergo multiple      
Compton scattering before they escape, hence the resulting spectrum is      
expected to be quasi-thermal irrespective of the details of the      
dissipation process.      
Following the dissipation, the plasma expands and cools. During the      
cooling phase, the plasma electrons are coupled to the photons via      
Compton scattering. The total energy in the photon component cannot      
exceed the total energy dissipated by the electrons at the dissipation      
phase, which is approximated by a fraction $\epsilon_e \leq 0.33$ of      
the total dissipated energy. As shown by \citet{PW04}, photons are      
expected to give up to $\sim 2/3$ of their energy to the plasma during      
the expansion phase that follows the dissipation.       
The plasma energy therefore increases by a factor smaller      
than $\sim 20\%$ of its initial energy during the expansion phase.

In addition to Compton scattering, possible sources of electrons    
energy loss during the expansion phase     
are synchrotron radiation and bremsstrahlung emission.  Synchrotron    
losses depend on the uncertain value of the magnetic field. An upper    
limit on this value during the dissipation phase can be found by    
assuming equipartition (a fraction $\epsilon_B \approx 0.33$ of the    
cocoon internal energy density is converted to magnetic field), which    
results in a value $B \simeq 10^{6} B_6$~G.  The characteristic momentum    
of electrons having Maxwellian distribution with temperature    
$\theta_{el} \ll 1$ is $\gamma \beta \simeq \theta_{el}^{1/2}$.  The    
cyclo-synchrotron power radiated by these electrons is therefore    
$P_{syn,0} = (4/9)[q^4 B^2 (\gamma \beta)^2]/(m_e^2 c^3) \simeq    
10^{-6} \, B_6^2 \, \theta_{el,-3} {\rm \, erg \, s^{-1}}$, where    
$\theta_{el} = 10^{-3} \theta_{el,-3}$ is used (see    
eq. \ref{eq:temp}).  The magnetic field, though, decreases quickly    
during the expansion phase.  Denoting by $R(t)$ the characteristic    
cocoon radius at time $t$, the number and energy density of electrons    
decrease as $n_{el}(t), u_{el}(t) \propto R(t)^{-3}$, thus $B(t)    
\propto R(t)^{-3/2}$ at times $t>t_0$ , where $t_0$ marks the    
beginning of the expansion phase.  For $t \gg t_0$, $R(t) \simeq c_s    
(t-t_0)$ where $c_s$ is the speed of sound (see below), therefore    
$P_{syn} (t) \propto t^{-3}$.  We thus conclude, that shortly after    
the beginning of the expansion, at $t \gtrsim (\Delta r_c / c_s)    
(P_{syn,0} t_{dyn} / m_e c^2)^{1/3}$ the electron energy loss time    
becomes longer than the dynamical time, which is comparable to the    
comoving diffusion time $t_{dyn} \simeq \Delta t_D^{co.}$. As a    
result, synchrotron losses do not affect the dynamics of the plasma    
during the expansion phase.

A similar conclusion can be drawn for bremsstrahlung emission.    
The rate of energy loss of electron by bremsstrahlung radiation    
during the dissipation is    
given by \citep[see, e.g.,][]{Rybicki79} $P_{brem} = (2 \pi^3    
\theta_{el}/3)^{1/2} (2^5 q^6 / 3 h m_e c^2) n_{el} \bar{g}_B \simeq 1.3    
\times 10^{-9} \, \theta_{el,-3}^{1/2} \, n_{el, 14.5} \, {\rm \, erg    
  \, s^{-1}} $, where $n_{el} = n_p(c) = 10^{14.5} n_{el, 14.5} \,    
\rm{cm^{-3}}$ is the electrons number density at the dissipation, and    
$\bar{g}_B$ is the     
averaged Gaunt factor. Since the electrons number density decreases    
quickly with     
time, we conclude that bremsstrahlung losses are insignificant.        
We can therefore approximate the expansion as adiabatic.

During the expansion phase, the expansion velocity of a plasma element      
(in its comoving frame) depends on the radial position of the      
element. The surface of the plasma expands at the velocity of sound,      
while internal parts expand at lower velocity, $v(r) \propto r$. The      
coupling of the expansion velocity to the position of a plasma element      
significantly complicates the problem\footnote{\citet{ST80} calculated      
analytically the light curve of photons escaping from a steady      
(non-expanding) plasma in the limit  $\tau_{\gamma e} \gg 1$, by      
solving the diffusion equation for photons. This method, however      
cannot be implemented here because of the coupling between the      
comoving plasma radial velocity and the radius of the plasma, $v(r)      
\propto r$.}.         
In order to calculate the emergence light curve and spectra, we      
therefore use a numerical code, based on Monte-Carlo      
method \footnote{We use a version of the Monte-Carlo code used in      
  \citet{PW04} for the calculation of photons energy loss during the      
  adiabatic expansion.}.

\subsection {The numerical model}      
\label{numeric}      
      
We consider a three-dimensional uniform, homogeneous plasma ball that      
expands in its rest frame at radial velocity $v(r) = [r/R(t)] c_s$,      
where $c_s= c/\sqrt{3}$ is the speed of sound, and $R(t) = R_0 + c_s      
t$ is the ball radius at comoving time $t$ after the dissipation (for      
convenience, we assume that the dissipation phase ends, and the      
adiabatic expansion begins at $t=0$. Thus, $R_0$ is the comoving      
plasma width at the end of the dissipation phase).  The plasma is      
characterized by an initial optical depth to scattering $\tau_{\gamma      
e}(t=0) = \tau_0$. As a result of the expansion, the number density of      
particles inside the plasma decreases as $n_{el} \propto R(t)^{-3}$, and      
the optical depth decreases as $\tau_{\gamma_e} (t) \propto R(t)      
n_{el}(t) \propto R(t)^{-2}$.      
The particle inside the plasma assume a Maxwellian distribution with      
initial temperature given by equation \ref{eq:temp}.     
As discussed above, particle cooling during the expansion is    
insignificant compared to adiabatic cooling. Therefore, at $t>0$, the     
particles cool adiabatically, hence their temperature decreases as      
$\theta_{el}(t) \propto R(t)^{-2}$.      
      
Since during the dissipation phase energy is dissipated inside the  
entire comoving width of the cocoon, we assume here that photons are  
injected uniformly inside the ball at $t=0$.  We consider two types of  
energy distributions of the injected photons: (a) mono-energetic  
injection, with initial (comoving) energy $\varepsilon_0 = (k_B  
T'(r_i)/m_e c^2)$, where $T'(r_i)$ is the electrons temperature at the  
dissipation radius, given by equation \ref{eq:temp0}; and (b) power  
law spectrum of the injected photons above $\varepsilon_0$, with a  
power law index $dn_\gamma/d\varepsilon \propto \varepsilon^{-2}$.

The characteristic energy of synchrotron photons emitted during the  
expansion phase is at least seven orders of magnitude lower than the  
plasma energy ($\varepsilon_{syn} \lesssim 10^{-5}$~eV for $B\sim  
10^6$~G), therefore these photons are neglected. Similarly, the power  
radiated in bremsstrahlung emission during the expansion is at least  
three orders of magnitude lower than the emitted power during the  
dissipation phase. We therefore neglect in the numerical calculations  
additional sources of radiation during the expansion phase, and  
consider Compton (and inverse Compton) scattering only during this  
phase.

Photons interact with electrons inside the plasma ball via Compton      
scattering (no absorption is considered). In the calculation, the      
exact differential Compton cross section is used.  For each escaping      
photon, the code traces its energy and the delay of the photon escape      
time compared to a hypothetical photon that was injected at the plasma      
center at $t=0$ and did not suffer any scattering before escaping the      
plasma.

\subsection{Numerical results}      
      
\begin{figure}                
\plotone{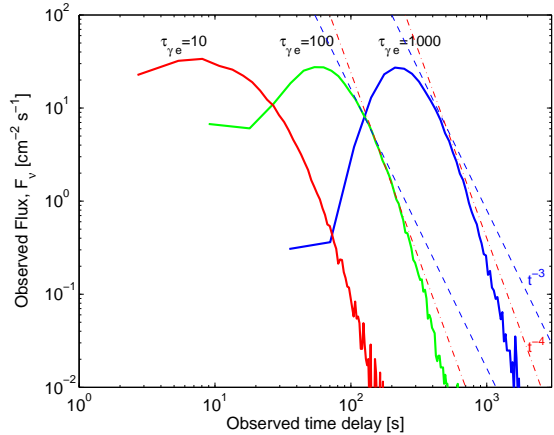}                
\caption{Light curves of photons emerging from an expanding cocoon,      
  characterized by initial optical depths $\tau_{\gamma e,c} = 10, 100,      
  1000$. Initial comoving radius $R_0 = 10^{12}$~cm, comoving      
  temperature $\theta_{el} = 10^{-3}$ injected photon energy      
  $\varepsilon_0 = 10^{-3}$ (in unites of $m_e c^2$), characteristic      
  Lorentz factor $\Gamma_c = 10$ and redshift $z=1$ are      
  assumed. The dissipated cocoon isotropic equivalent energy  
  $E_{c,52} = 0.3,1,3$ respectively is considered. 
Photons are injected uniformly inside the sphere.       
Similar light curves are obtained for photon injection with power law      
  energy distribution above $\varepsilon_0$.       
 Each of the presented curves is produced by $10^5$ Monte-Carlo runs.  }      
\label{fig:lightcurve}                
\end{figure}

The light curves obtained for three different values of the optical    
depth, $\tau_{\gamma e} = 10^1, 10^2, 10^3$, and comoving width $R_0    
= 10^{12}$~cm in all three cases are presented in Figure    
\ref{fig:lightcurve}. The cocoon isotropic equivalent dissipated energy  
is $E_{c,52} = 0.3, 1, 3$ respectively, and we assume a redshift $z=1$ in a 
flat universe for calculating the observed flux. These values correspond to  
a kinetic energy dissipation phase that occurs at radii $\approx 10^{13}$~cm  
  (see eq. \ref{eq:tau}).      
The mean time delays of photons in the three    
presented graphs are $8, 60, 210$~s for $\tau_{\gamma e} = 10^1, 10^2,    
10^3$ respectively. These values are in very good agreement with the    
analytical estimate of equation \ref{eq:t_diffuse}.  Clearly, the    
light curves at late times decrease according to $F_\nu(t) \propto    
t^{-\alpha}$ with $\alpha \approx 3-4$, irrespective of the exact    
value of the optical depth.  Similar steep decrease in the light    
curves at late times was found by \citet{ST80} for the case of    
non-expanding plasma and various geometries considered there. This    
suggests that a steep decrease in the light curve is a general    
property of emission from plasma characterized by number of scattering    
higher than few tens \citep[in an expanding plasma, the mean number of    
scattering of a photon before it escapes is $n_{sc.} \propto    
\tau_{\gamma e}$, while in a steady plasma $n_{sc.} \propto    
\tau_{\gamma e}^2$; see, e.g.][]{PW04}.

\begin{figure}                
\plotone{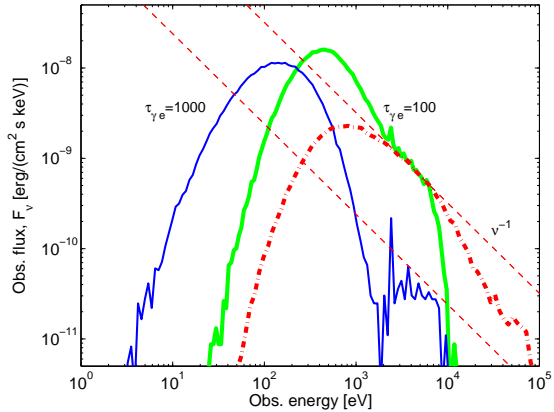}                
\caption{Resulting spectra of the emerging photons for the cases of      
  $\tau_{\gamma e,c} = 100, 1000$. Solid lines present injection of    
  mono energetic photons at energy $\varepsilon_0 = 10^{-3}$ (in units    
  of $m_ec^2$), and dash-dotted line presents injection of photons       
  having power law energy distribution with power law index $p=2$       
  above $\varepsilon_0$, as might be expected from synchrotron    
  emission above the peak and cooling frequencies. Thick lines are for    
  $\tau_{\gamma e} = 100$, and the thin line presents the case of    
  $\tau_{\gamma e} = 1000$. All other parameters are the same as in    
  figure \ref{fig:lightcurve}. For $\tau_{\gamma e} = 100$, the    
  spectral        
  slope in the XRT range ($0.3-10 \keV$) is close to $\beta = 1$,    
  while for $\tau_{\gamma e} = 1000$    
  a single spectral slope is marginally consistent with this result.  }      
\label{fig:spectrum}                
\end{figure}

The emerging photons' spectra are presented in Figure \ref{fig:spectrum}.    
The photon injection is approximated as monoenergetical, with    
energy equal to the peak energy of the thermal distribution of    
the electrons, $\varepsilon_0 = \theta_{el} m_e c^2 = 500 \eV$ (in the    
plasma frame). The observed energy of the injected photons is    
therefore $\varepsilon_0^{ob.} = \varepsilon_0 \times \Gamma_c /(1+z)    
= 2.5 \keV$, where $\Gamma_c = 10$ and redshift $z=1$ assumed. For    
high value of the optical depth, most of the photons lose energy    
before they escape. As shown in \citet{PW04}, for $\tau_{\gamma e} =    
100$ about 70\% of the energy lost by the photons is transferred to    
the bulk motion of the expanding plasma, and the remaining is    
transferred to thermal energy of the electrons.    
This energy loss results in the quasi-thermal spectrum obtained at low    
energies.  Still, for a uniform injection of photons as is considered    
here, the high energy tail of the emerging photon spectra is a    
power-law like with power law index $\beta \simeq 1$ ($F_\nu \propto    
\nu^{-\beta}$), in the energy range $0.3-10 \keV$, which is the {\it    
Swift} XRT energy band. The case $\tau_{\gamma e} = 1000$    
presented in figure \ref{fig:spectrum}, shows an upper limit on the    
cocoon optical depth that is still consistent with the    
observations. Due to the large number of scattering, the photon index    
in the {\it Swift} XRT range is not well defined, and the data is    
marginally consistent with a power law index $\beta \simeq 1$ in this    
range.     
    
A similar result is obtained when one considers the more      
realistic case of power law photon injection above the characteristic      
energy $\varepsilon_0^{ob.}$, as is expected, e.g., if the main      
radiative mechanism is synchrotron radiation, or alternatively,      
multiple Compton scattering of a photospheric component      
\citep{PMR05b}. In figure \ref{fig:spectrum}, the dash-dotted line      
presents the case of power law injection of photons above      
$\varepsilon_0^{ob.}$, with power law index $p=2$. In this case as      
well, the obtained spectral slopes have power law index $\beta \simeq 1$      
in the {\it Swift} XRT energy band.

\section{Comparison with data: the cases of GRB050315 and GRB050421}      
\label{sec:data}      
      
\begin{figure}                
\plotone{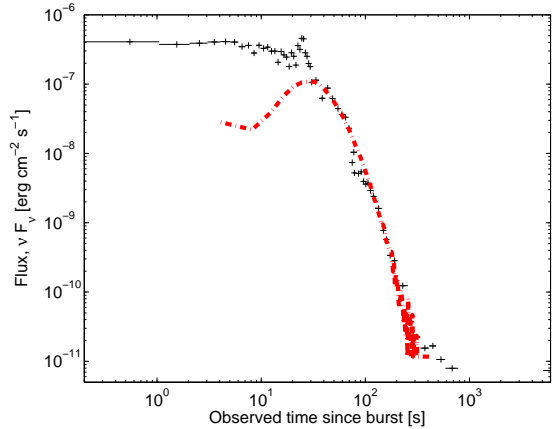}                
\caption{Fitting of the decline part in the XRT data of GRB050315      
  with the results of the numerical simulation light curve. Parameters      
  used in the fitting are $\tau_{\gamma e,c} = 100$, $R_0 = 10^{12}$~cm,      
  $\Gamma_c = 30$, $E_{c,52} = 2.5$ and $z=1.949$.  }      
\label{fig:data1}                
\end{figure}

\begin{figure}                
\plotone{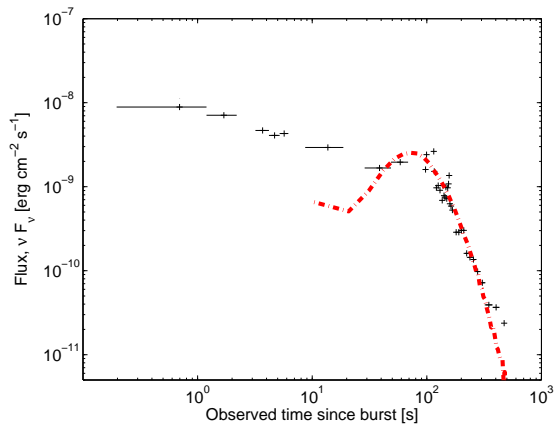}                
\caption{Fitting of the decline part in the XRT data of GRB050421       
with the results of the numerical simulation light curve.      
Here, $\Gamma_c = 8$, $E_{c,52} = 0.1$ and $z=1$ were considered, and  
all other numerical free        
parameters have the same values as in figure \ref{fig:data1}.      
}                
\label{fig:data2}                
\end{figure}

The steep decline in the XRT light curves observed in many bursts      
\citep{Obrien05} is similar to the steep decline in the light curves      
obtained by the Monte Carlo simulation. We show in Figures      
\ref{fig:data1},\ref{fig:data2} two fits of the data of GRB050315 and      
GRB050421 obtained from P. Obrien by the numerical model results.      
These two bursts did not show significant flaring activity, and had a      
good time coverage during the steep decay phase of the emission.      
       
In fitting the data we assumed a cocoon optical depth at the
dissipation radius $\tau_{\gamma e,c} = 100$. This value is consistent
with the assumption that the dissipation radius is $\gtrsim
10^{13}$~cm.  The redshift of GRB050315 is $z=1.949$
\citep{Vaughan05}, and the redshift of GRB050421 is unknown, thus is
taken here to be $z=1$.  The values of the cocoon Lorentz factors
during the cocoon dissipation phase $\Gamma_c$ were determined using
the observed time delays at the beginning of the decay. The resulting
Lorentz factors, $\Gamma_c = 30, 8$ in the two cases are of the same
order of magnitude, $\Gamma_c \sim 10$.  The value of the dissipated
cocoon energy required to fit the data depends on the assumption about
the injected photon distribution. In the case of GRB050315, an
isotropic equivalent cocoon energy $E_{c,52}=1.5$ ($E_{c,52}=2.5$) is
required in the monoenergetic (power law injection) scenario, while in
the case of GRB050421 cocoon energy $E_{c,52}=0.03$ ($E_{c,52}=0.1$)
is required in these two scenarios.  The values found are consistent
with the order-of-magnitude estimates presented in
\S\ref{sec:time_delay} based on theoretical arguments.

The data of GRB050421 (figure \ref{fig:data2}) shows clear distinction    
between the early shallow decay observed at the first $\sim 30$~s,    
which is attributed to prompt emission, and the rise of the flux    
followed by a steep fall at $\sim 100$~s which is attributed in our    
model to emission from the cocoon. This time separation is less    
significant in the case of GRB050315 presented in figure    
\ref{fig:data1}, where a flare at $\sim 25$~s precedes the cocoon    
emission, which peaks at $\sim 30$~s. We further discuss this time    
separation issue in \S\ref{sec:summary} below.     
    
\begin{figure}                
\plotone{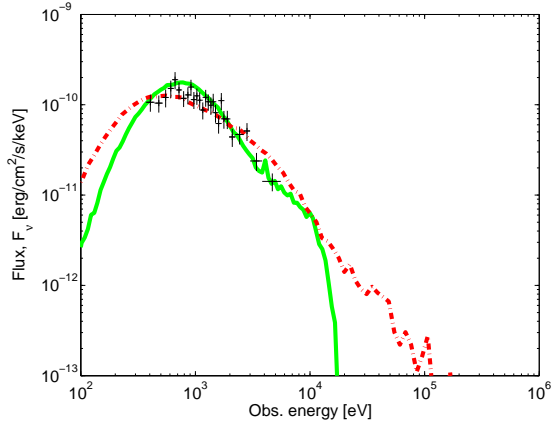}                
\caption{Fitting of the XRT spectrum of GRB050315      
  with the results of the numerical simulation light curves. Parameters      
  used in the fitting are $\tau_{\gamma e,c} = 100$, $R_0 = 10^{12}$~cm,      
  $\Gamma_c = 30$ and $z=1.949$.       
  The cocoon energy and comoving temperature are $E_{c,52}=1.5$,  
  $\theta_{el} = 8.3\times 10^{-4}$       
 for the mono-energetic (solid line),  $E_{c,52}=2.5$, $\theta_{el}=  
  3.3\times 10^{-4}$ for the  power law (dash dotted line) photon      
  injection assumptions. }      
\label{fig:spectra1}                
\end{figure}

\begin{figure}                
\plotone{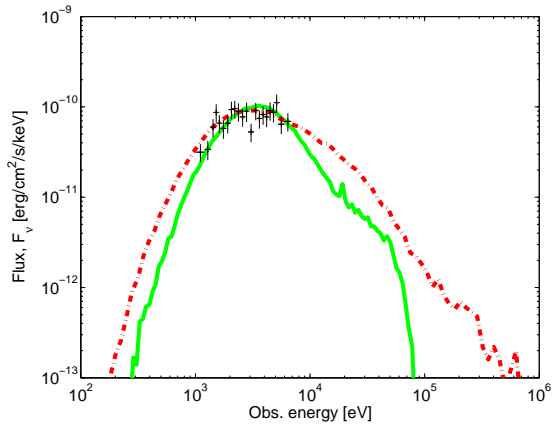}                
\caption{Fitting of the XRT spectrum of GRB050421      
  with the results of the numerical simulation light curves. Parameters      
  used in the fitting are $\tau_{\gamma e,c} = 100$, $R_0 = 10^{12}$~cm,      
  $\Gamma_c = 8$ and $z=1$.       
  The cocoon energy and comoving temperature are $E_{c,52}=0.03$,  
  $\theta_{el} = 10^{-2}$ for the mono-energetic (solid line),  
  $E_{c,52}=0.1$, $\theta_{el}= 5\times 10^{-3}$ for the  power law  
  (dash dotted line) photon injection assumptions. }      
\label{fig:spectra2}                
\end{figure}

Fitting of the XRT spectra in these two cases are presented in Figures  
\ref{fig:spectra1}, \ref{fig:spectra2}. In the cases presented, we use  
the same parameters of the fits presented in figures  
\ref{fig:data1},\ref{fig:data2} with a re-normalization of the flux to  
  the cocoon emission time.  We assume initial normalized  
electrons temperature $\theta_{el} = 10^{-3}$ and injected photon  
energy $\varepsilon_0 = \theta_{el} m_e c^2 = 500 \eV$ in the  
monoenergetic injection case. In the power law photon injection  
assumption, a power law index $p=2$ above $\varepsilon_0$ in the  
photon energy distribution is assumed.  
    
In fitting the data of GRB050315, the numerical results were shifted    
by a factor of 1.2 (3) in the mono-energetic (power law) injection    
cases.  A physical origin of these shifts is that the temperature of    
the cocoon is different than the assumed numerical value of    
$\theta_{el} = 10^{-3}$ by equivalent factors. We can thus summarize    
that the comoving temperature of the cocoon is $\theta_{el} =    
8.3\times 10^{-4}$ ($3.3 \times 10^{-4}$) for the monoenergetic (power    
law) injections assumptions used. Similarly, the comoving temperature    
of the plasma for the case of GRB050421 is $\theta_{el} = 10^{-2}$    
($5\times 10^{-3}$) for monoenergetic (power law) photon injection    
cases.

For GRB050315, the spectra of the BAT data at late times is fitted      
with the XRT data with a single power law \citep{Vaughan05}. This is      
consistent with the numerical results of the light curve presented in      
Figure \ref{fig:spectra1} for the more realistic scenario of power law      
injection of photons. BAT data of GRB050421 exists only at the first      
$\sim 100$~s, before the steep decay phase of the light curve. We thus      
conclude that the BAT photons originate from the jet, and not from the      
cocoon plasma.          
      
The two fits presented here are only representative cases. Similar    
fits can be obtained for many light curves in \citet{Obrien05} sample,    
which show power law indices during the decay phase in the range    
$\alpha \simeq 2 - 4$.  The similarities found between the numerical    
model light curves for different values of the optical depth imply    
that the data can be fitted with various values of the free parameters    
$\tau_{\gamma e,c}$ and $\Gamma_c$. However, the demand that    
$\Gamma_c$ is of the order of few tens, and not larger than $\sim    
100$, and the demand that the spectrum differs from a black-body    
restricts the optical depth in both cases considered here to be    
$\tau_{\gamma e,c} \sim 100$.    
    
The results presented here thus show that using values of the optical  
depth and cocoon temperature close to the ``canonical'' values  
considered in \S\ref{sec:time_delay} (see eqs. \ref{eq:tau},  
\ref{eq:temp}), emission from the cocoon can account for the observed  
steep decay spectra. The injected photons are assumed in all cases to  
have monoenergetic distribution at energy similar to the peak energy  
of the thermal electrons inside the cocoon, or to have a power law  
distribution above this energy. These photons therefore originate by  
dissipation mechanism inside the cocoon, without any assumptions on  
photons produced by the prompt emission.  For assumed cocoon optical  
depth, by fitting the model to the data one can determine the values  
of the cocoon Lorentz factor $\Gamma_c$, the electrons temperature  
$\theta_{el}$ and the dissipated cocoon energy, $E_c$.

\section{Summary and discussion}      
\label{sec:summary}      
      
In this paper, we have considered emission from an expanding cocoon as      
a possible explanation to the steep decay in the {\it Swift} XRT light      
curve observed in many bursts after few tens - few hundreds of      
seconds.  We showed in \S\ref{sec:time_delay} that due to the high      
optical depth in the cocoon, photons produced inside the cocoon at      
similar radii to the estimated dissipation radius of the kinetic      
energy of the jet, are observed after a diffusion time delay of few      
hundreds of seconds. We then presented in \S\ref{sec:light_curve} our      
numerical code, and showed the emergent photons light curve and      
spectra. We showed that the emergent photons light curves and spectra      
are similar to the decaying part of the light curves and the XRT      
spectra  observed in GRB050315 and GRB050421,       
and argued that due to the similarities in many of the {\it Swift} XRT      
light curves, the numerical results fit the data in many bursts.      
      
Due to the complexity of the problem, we did not obtain a simple      
analytical formulae for the light curves. However, we showed that the      
numerical results can be approximated as a power law over a limited      
time interval and energy range, which is similar to the energy range      
observed by the {\it Swift} XRT.       
Due to the large number of scattering the photons undergo before they      
escape, the resulting spectrum in the XRT band is by large independent      
on the initial photons spectrum, which by itself      
depends on the details of the dissipation process.      
We therefore conclude that observations of the spectrum during the      
decay phase are not expected to unveil much of the mystery of the      
dissipation mechanism, if the cocoon model is correct.       
       
We found that the shape of the light curves are similar for different      
values of the optical depth (figure \ref{fig:lightcurve}).      
Light curves similar to the light curves presented here were also      
obtained       
by \citet{ST80}, for the case of non-expanding plasma. These facts      
suggest that the shape of the light curve may be common to many      
astrophysical phenomena in which photons undergo tens to hundreds of      
scattering before they escape the plasma in which they are produced.      
This may also explain the similarities found in the shape of the light      
curves in many different bursts in the sample of \citet{Obrien05}.      
      
These similarities also imply that the data could be fitted with  
various values of the free parameters - $\tau_{\gamma e,c}$,  
$\Gamma_c$ and the comoving temperature $\theta_{el}$. We showed in  
\S\ref{sec:data} two fits for the light curves and energy spectra of  
two different bursts, assuming that photons are produced at the  
dissipation inside the cocoon, without additional prompt emission  
photon source. We found that the fitted values of the cocoon Lorentz  
factor $\Gamma_c$, the cocoon energy $E_c$ and the electrons  
temperature $\theta_{el}$ are   
close to their ``canonical'' values calculated in  
\S\ref{sec:time_delay} based on theoretical arguments.  Restrictions  
on the values of these parameters are obtained from the physical  
demands that $\Gamma_c$ is of the order of few tens, and not larger  
than $\sim 100$, and that the plasma temperature should not exceed the  
theoretical values predicted in \S\ref{sec:time_delay}.  
      
The sample of Swift-XRT light curves presented in Figure 3 of  
\citet{Obrien05} shows significant variety. In many cases (e.g.,  
GRB050319, GRB050412, GRB050422, GRB050713A, GRB050803 and many more),  
there is a clear distinction between the early (prompt) emission  
phase, and the steep decay phase. Following an early shallow decline,  
the flux rises before entering the steep decay phase. This pattern can  
be attributed to emission from the cocoon (see figure  
\ref{fig:lightcurve}). In other cases (e.g., GRB050713B, GRB050814 and  
GRB 050819) a steep fall in the flux follows the end of the prompt  
emission, without a significant rise preceding the decay. In other  
cases, a shallow decay of the flux is observed over hundreds -  
thousands of seconds.

The model presented here suggests that (at least part of) these    
observations are the result of emission from the cocoon. The extension    
of the prompt emission to tens - hundreds of seconds implies that in    
many cases the emission from the cocoon is partially obscured by the    
prompt emission. This may explain the lack of rise in the flux in many    
cases which show steep decay. Obviously, the model presented here    
cannot account for the entire variety of the light curves observed.    
Nonetheless, the facts that rise in the flux was observed prior to the    
steep decay in several cases, and the steepness of the decay - $\alpha    
\approx 2.5 - 4$ in many cases, are difficult to account for in models    
that do not consider the existence of the cocoon.

In addition to the many cases of long bursts which show a steep  
decline in the XRT light curve, a similar behavior was observed in  
one case of a short burst - GRB050724. While the nature of short  
bursts progenitors is still not fully understood, it is most likely  
that these bursts do not originate from the explosion of massive stars,  
hence the cocoon model cannot be similarly motivated in this case. The decay  
observed in GRB050724 is so far unique, and was not observed in other  
short bursts. The present explanation, which applies to long bursts,  
does not preclude a high-latitude emission effect interpretation in  
some bursts, e.g., in short bursts, or in some fraction of long  
bursts.    
     
Our model assumes an unspecified dissipation mechanism that converts     
cocoon kinetic energy into radiation at radii $\lesssim 10^{13}$ cm,     
below the radius where the cocoon becomes optically thin to scattering.     
This radius could be as low as the stellar surface $r\sim 10^{11}$ cm,     
but could be anywhere up to the photosphere.     
Such a dissipation may be caused by magnetic reconnection      
\citep{GS05}, interaction with the expanding envelope of the star      
\citep{MR01} or extended central engine activity and refreshed shocks      
in the jet that could interact laterally with the cocoon region, and      
may derive oblique shocks into the cocoon.      
 As presented in      
\S\ref{sec:appendix2}, similar time delays would in principle occur if      
the cocoon photons were advected from the core of the star, without      
any dissipation of the cocoon kinetic energy at larger      
radii. However, in this latter case the very high optical depth at      
$\sim 10^{11}$~cm would imply that the emergent spectrum is very      
close to black-body, which for the bursts discussed here is not     
consistent with the observations.

In contrast to the internal collision model, which predicts multiple      
dissipation phases resulting in long duration prompt emission, the      
cocoon model predicts only a single dissipation phase of the cocoon      
plasma. This model therefore disfavors a second phase of steep decay.      
The flares that are observed in many bursts thus are predicted to have      
a different source, e.g., extended central engine activity.       
Nonetheless, as discussed above, if the dissipation of the jet kinetic      
energy occurs at small enough radii where the optical depth is higher      
than a few tens, similar light curves are expected.

\acknowledgements     
We wish to thank Paul O'Brien and Dave N. Burrows for useful    
discussions, and Paul O'Brien and Kim Page for providing us the data    
of GRB050315 and GRB050421. AP wishes to thank Ralph A.M.J. Wijers and    
Ed van den Heuvel for useful discussions.  This research was supported    
by NWO grant 639.043.302, by the EU under RTN grant HPRN-CT-2002-00294    
and by NSF AST0307376, NASA NAG5- 13286 and NSF PHY99-07949.    
     
\appendix      
\section{The diffusive time delay in an expanding plasma}      
\label{sec:appendix}      
      
Assuming that following the dissipation phase the cocoon plasma      
expands at velocity close to its asymptotic velocity, $c_s =      
c/\sqrt{3}$ the mean free path of photons increases with comoving time      
as $l(t) =l_0 (1 + v t/\Delta r)^3$, where $l_0 = \Delta r_c/      
\tau_{\gamma e,c}$ is the mean free path at the end of the dissipation      
phase, and constant expansion velocity $v=c_s$ assumed.           
 For a large number of scattering, $\tau_{\gamma e,c} \gg 1$, the      
 distance traveled by a photon before it escapes is      
      
\beq      
c \Delta t_d^{co.} = \int_0^{\Delta t_d^{co.}} {dl \over dt} dt =       
l_0 \left( 1 + {v \Delta t_d^{co.} \over \Delta r_c} \right)^3       
\approx {\Delta r_c \over \tau_{\gamma e,c}}       
\left({v \Delta t_d^{co.} \over \Delta r_c} \right)^3,      
\eeq      
where $\Delta t_d^{co.}$ is the comoving photon diffusion time.       
We thus find that in the observed frame, the diffusion time is given by      
\beq      
\ba{lll}      
\Delta t_d & = & {\Delta t_d^{co.} \over \Gamma_c} \approx \sqrt{{c \over      
    v^3}} {\sqrt{\tau_{\gamma e,c}} \Delta r_c \over \Gamma_c}      
\simeq 3^{3/4} {\sqrt{\tau_{\gamma e,c}}      
\Delta r_c \over c \Gamma_c} \nonumber \\      
& = & 250 \, E_{c,52.5}^{1/2} r_{\star,11}^{3/2}      
r_{i,13}^{-3/2} \Gamma_{c,1}^{1/2} \, \rm{s}.      
\ea      
\label{eq:t_diffuse}      
\eeq      
The diffusive time delay calculated above is of the order of few      
hundreds of seconds for the assumed values of the free parameters.      
This time delay is much longer than the escape time delay of the      
cocoon plasma from the core of the star $\Delta t_{esc}$, and it is longer     
than the geometrical time delay $\Delta t_{g,jet}$ of the prompt radiation     
from the jet, but it is comparable to the geometrical time delay from the     
photosphere of the cocoon, $\Delta t_{g,c}$, as long      
as the optical depth to scattering inside the cocoon is high.

\section{Alternative approaches for the calculation of the diffusion      
  time delay}      
\label{sec:appendix2}      
      
Photons that are produced in an expanding plasma in regions of high      
optical depth (i.e., at dissipation radii $r_D \ll r_{ph}$) emerge      
from it as the plasma reaches the photospheric radius, $r_{ph}$.      
The observed time delay of these photons compared to photons that did      
not suffer any scattering can be calculated in two      
alternative ways:      
(a) calculation of the diffusion time delay in an expanding plasma, as      
was done here; and       
(b) since these photons are advected outward with the flow from       
the emission radius $r_D$ to the photospheric radius, $r_{ph}$, the      
time delay should be equal to the ``geometrical'' time delay of      
photons emerging from $r_{ph}$.      
Here, we show that the two different approaches lead to the same      
result,    
provided the optical depth at the dissipation radius is     
high and that the plasma expands at velocity close to $c$.      
        
In order to include a variety of possible dissipation mechanisms (such      
as, e.g., magnetic reconnection),       
we assume that the comoving width of the plasma at the dissipation      
radius is arbitrary, and denoted by $\Delta r_D$.      
At larger radii  $r>r_D$ the dissipated region is      
advected with the flow, therefore the width of this region increases      
with radius, $\Delta r \propto r$.      
For a single (not continuous) energy release, the number density of      
protons at radius $r$ is $n_p(r) = E/(4 \pi r^3 m_p c^2)$,      
thus the optical depth at radius $r$ is       
\beq      
\tau(r) = { E \sigma_T \Delta r_D \over 4 \pi r_D r^2 m_p c^2 }      
\eeq      
where $\Delta r (r) = \Delta r_D (r/r_D)$.       
The photospheric radius is given by $r_{ph} = r(\tau=1)$, or      
\beq      
r_{ph} = \left( { E \sigma_T \Delta r_D \over 4 \pi r_D m_p c^2 }\right)^{1/2}.      
\eeq       
The diffusive time delay is approximated by       
\beq      
\Delta t_D^{ob.} \simeq {\sqrt{\tau(r=r_D)} \Delta r_D \over \Gamma c}      
= {r_{ph} \over \Gamma c} {\Delta r_D \over r_D} = {\Delta r(r_{ph}) \over      
  \Gamma c}.        
\eeq      
We thus find that the diffusive time delay is similar to the time    
delay resulting from the finite width of the plasma at the    
photospheric radius \citep[see][]{W97c}. In addition, photons that are    
emitted at the photospheric radius from a point not on the line of    
sight suffer a time delay $\sim r_{ph}/\Gamma^2 c$, which is thus the    
minimum observed time delay.  For the specific case of dissipation by    
internal shock waves, the comoving width is $\Delta r = r/\Gamma$, the    
time delay due to the finite width of the plasma is similar to the    
time delay due to emission from positions off the line of sight, and    
we obtain the familiar result $\Delta t_D^{ob.} \simeq r_{ph}/\Gamma^2    
c$.  However, the above analysis shows that in the more general case    
of arbitrary plasma width $\Delta r > r/\Gamma$, the diffusive time    
delay is equal to the time delay due to the finite width of the    
plasma, which determines the observed time delay.

\end{document}